\documentclass{egpubl}
\usepackage{eg2026}
 
\ConferencePaper        
\CGFStandardLicense

\usepackage[T1]{fontenc}
\usepackage{dfadobe}  

\biberVersion
\BibtexOrBiblatex
\usepackage[backend=biber,bibstyle=EG,citestyle=alphabetic,backref=true]{biblatex} 
\electronicVersion
\PrintedOrElectronic

\ifpdf \usepackage[pdftex]{graphicx} \pdfcompresslevel=9
\else \usepackage[dvips]{graphicx} \fi

\usepackage{egweblnk} 

\usepackage{amsfonts}
\usepackage{amsmath}
\usepackage{xcolor} 

\title[Physics-Based Motion Tracking of Contact-Rich Interacting Characters]{Physics-Based Motion Tracking of Contact-Rich Interacting Characters}

\author[X. Zhang \& Z. Chang \& Q. Men \& H. P. H. Shum]
{\parbox{\textwidth}{\centering Xiaotang Zhang$^{1}$\orcid{0000-0003-0822-9064}, Ziyi Chang$^{1}$\orcid{0000-0003-0746-6826}, Qianhui Men$^{2}$\orcid{0000-0002-0059-5484}, and Hubert P. H. Shum\thanks{Corresponding author}$^{1}$\orcid{0000-0001-5651-6039}}
        \\
{\parbox{\textwidth}{\centering $^1$Durham University, United Kingdom\\
        \{xiaotang.zhang, ziyi.chang, hubert.shum\}@durham.ac.uk \\
         $^2$University of Bristol, United Kingdom\\
         qianhui.men@bristol.ac.uk\\
       }
}
}

\begin{document}

\teaser{
 \includegraphics[width=0.9\linewidth]{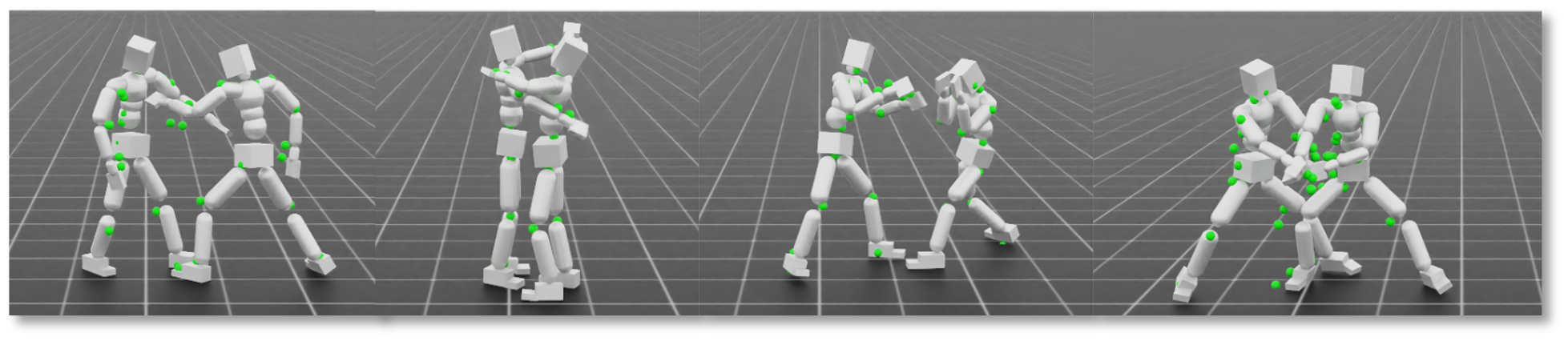}
 \centering
  \caption{Physics-based motion tracking of two humanoid characters performing contact-rich interactions such as boxing, pushing, and grappling. The objective is to track and reproduce stable motions under frequent physical contacts and complex force exchanges.}
\label{fig:teaser}
}

\maketitle
\begin{abstract}
Motion tracking has been an important technique for imitating human-like movement from large-scale datasets in physics-based motion synthesis. However, existing approaches focus on tracking either single character or a particular type of interaction, limiting their ability to handle contact-rich interactions. Extending single-character tracking approaches suffers from the instability due to the challenge of forces transferred through contacts. Contact-rich interactions requires levels of control, which places much greater demands on model capacity.
To this end, we propose a robust tracking method based on progressive neural network (PNN) where multiple experts are specialized in learning skills of various difficulties. Our method learns to assign training samples to experts automatically without requiring manually scheduling. 
Both qualitative and quantitative results show that our method delivers more stable motion tracking in densely interactive movements while enabling more efficient model training.



\end{abstract}  

\section{Introduction}

Physics-based motion tracking enables the synthesis of physically valid movements by using the next-frame pose as a control signal. Despite extensive research, existing approaches focus exclusively on  either single-character motion tracking \cite{peng2018deepmimic,luo2023perpetual}, or a particular type of interactions with a task-specific controller \cite{won2021control,zhu2023neural,luo2024smplolympics}, limiting their ability to handle contact-rich interactions.

Extending single-character motion tracking methods to interaction settings faces a fundamental challenge: preserving stability while transferring forces through contacts. In physical simulation, identical poses may have different underlying contact forces that may vary significantly. When characters are controlled independently by single-character tracking methods without modeling force transfer, contact-rich interactions become unstable and may even fail to track, as one character may receive inconsistent and uncontrolled forces from its opponent. For example, different poses in highfive have different distributions of contact forces between hands. Single-character tracking is unable to handle this since it does not take potential transferred forces into account. When sudden forces are transmitted through contacts, they can lead to oscillations or unmeaningful movements. 



Since contact-rich interactions like boxing or dancing introduce physical perturbations, another challenge of this task is that they places much greater demands on model capacity. This added complexity significantly increases the risk of catastrophic forgetting in tracking network \cite{luo2023perpetual}. Prior work on physics-based humanoid interactions \cite{zhu2023neural,won2021control,luo2024smplolympics} usually addresses this by restricting the problem scope: training skill-specific policies \cite{won2021control,peng2018deepmimic}, adopting multi-stage learning paradigms \cite{zhu2023neural}, or focusing on sparse and long-range interactions \cite{luo2024smplolympics}. However, these strategies either incur high training costs or compromise performance on the dense, contact-rich interaction details. 

To overcome these challenges, we propose a progressive, all-in-one mixture-of-experts architecture. 
Intuitively, interactions naturally involve different levels of control complexity. For example, in boxing, low-level control maintains stable locomotion such as stance and stepping, while mid-level to high-level control governs reactive behaviors such as dodging and blocking. These controls operate hierarchically, with higher-level strategies relying on the stability of low-level control. 
Building on this intuition, our method eliminates the need for multi-stage skill-specific training \cite{luo2023perpetual,peng2018deepmimic} by introducing a progressive mixture-of-experts design, enabling a single policy for contact-rich interactions. 

Specifically, we introduce a progressive training strategy for the mixture-of-experts architecture, drawing inspiration from the progressive neural network (PNN) framework \cite{rusu2016progressive}. This strategy addresses the instability in robot states caused by dense contacts in contact-rich interactions. By routing samples based on tracking error, the system automatically assigns contact-free, high-reward, stable samples to base experts and contact-heavy, low-reward, unstable samples (where force transfer occurs and is challenging to track) to specialized experts. In contrast to the original PNN where expert policies are manually assigned to distinct datasets or tasks, our approach removes the need for such predefined partitioning. By training directly on the entire motion dataset, the framework autonomously allocates samples of varying difficulty to the most appropriate experts.

We demonstrate our method on InterHuman \cite{liang2024intergen}, achieving robust tracking accuracy and smooth transfer across different interaction patterns. We also validate the stability of our approach under perturbations introduced via obstacles or next-frame poses. Finally, we analyze the contributions of individual experts, highlighting the model’s ability to capture different levels of torque control. Our contributions are summarized as follows:
\begin{itemize}
\item \textit{All-in-one framework} We propose an architecture for physics-based motion tracking in contact-rich interactions, removing the reliance on task-specific or multi-stage controllers.
\item \textit{Progressive mixture-of-experts} We introduce a progressive mixture-of-experts strategy that gradually adding new experts to model the hierarchical levels of torque control, enabling stable prediction of joint torques.
\item \textit{Robust interaction tracking} We demonstrate the effectiveness of our method on large-scale datasets, showing superior tracking accuracy, smooth and realistic interactions across diverse tasks, and strong robustness under perturbations.
\end{itemize}

\section{Related Works}

\subsection{Physics-based Humanoid Motion Tracking}
Since no ground-truth data exist of human joint actuation and physics simulators are often non-differentiable, a policy, aka controller, is often trained to track and mimic human motion using deep reinforcement learning (RL). 
From \cite{peng2018deepmimic}, RL-based motion tracking has gone from imitating single clips to large-scale datasets \cite{chentanez2018physics,wang2020unicon,luo2023perpetual,fussell2021supertrack}. 
Among them, a mixture of experts \cite{won2020scalable}, differentiable simulation \cite{ren2023diffmimic}, and external forces \cite{yuan2020residual} have been used to improve the quality of motion imitation. 
Recently, Luo et al. \cite{luo2023perpetual} allows a single policy to mimic almost all of AMASS and recover from falls. 
Luo et al. \cite{luo2023universal} improve \cite{luo2023perpetual} to track all AMASS and distill its motor skills into a latent space. 
Luo et al. \cite{luo2024omnigrasp} focuses on humanoid motion imitation with articulated fingers. 
Luo et al. \cite{luo2024real} focuses on tracking whole body motions based on head mounted devices. 
Xu et al. \cite{xu2025parc} proposes a controller to reduce incorrect contacts or discontinuities for traversing new terrains. 
Tessler et al. \cite{tessler2024maskedmimic} learns a physics-based controller to provide an intuitive control interface without requiring tedious reward engineering for all behaviors of interest. 
Juravsky et al. \cite{juravsky2024superpadl} trains controllers on thousands of diverse motion clips via progressive expert distillation. 
CLoSD~\cite{tevet2025closd} uses a diffusion model~\cite{chang2026design} for tracking.
While existing tracking methods have advanced physics-based generation of single character, they cannot generalize into two-character interactions due to the force transfer and significantly increased complexity.

\subsection{Physics-based Two-Character Interaction Synthesis}
Although physics-based methods has shown promising results for individual characters performing a wide variety of behaviors, there exist only a few studies for multi-character animations. 
Park et al. \cite{park2019learning} shows an example of chicken hopping and fighting with pre-defined discrete actions as well as target goals. While it shows the potential of physics-based two-character interaction synthesis, their interactions are simple and sparse. 
Won et al. \cite{won2021control} presents a control strategy for two boxing or fencing characters by defining goal-oriented rewards. However, the control policy only demonstrates effectiveness on a specific interaction task with such task-specific rewards. 
Zhang et al. \cite{zhang2023simulation} proposes a new reward formulation to facilitate various types of spatially and temporally dense interactions for full-body humanoid characters. However, their controllers are imitation controllers that cannot perform interactions that do not exist in the reference motions. 
Zhu et al. \cite{zhu2023neural} combines discrete latents with reinforcement learning to synthesize two-character boxing motions. However, their policy requires multiple stages to train and their control is specific to tasks. 
Younes et al. \cite{younes2023maaip} leverages adversarial imitation learning to generalize the idea of motion imitation for one character to deal with both the interaction for two physics-based characters. Their control policy is also specific to an interaction type and cannot generalize to multiple types. 
Liu et al. \cite{liu2024physreaction} focuses on reactive motions of one character when it interacts with the other character. 
Luo et al. \cite{luo2024smplolympics} proposes a benchmark for simulating two-character interactions in the field of sports. However, their interactions are sparse and far from each other. 
Xie et al. \cite{xie2022learning} introduces a layer-wise mixture-of-experts architecture to integrate a diverse range of high-precision soccer juggling skills into a single physics-based character controller. Their system utilizes a task-description framework based on control graphs and success-adaptive random walks to facilitate the efficient learning of complex motor tasks and robust transitions between different body-part interactions.
While kinematics-based interaction modeling~\cite{zhang2025motion,zhang2025real,chang2025large} involve dense interactions, they do not consider physics. Our work focuses on generating contact-rich interactions within an all-in-one pipeline, which has been one of the challenging problems in physics-based two-character interaction animation.

\begin{figure*}[tbp]
  \centering
  \includegraphics[width=\linewidth]{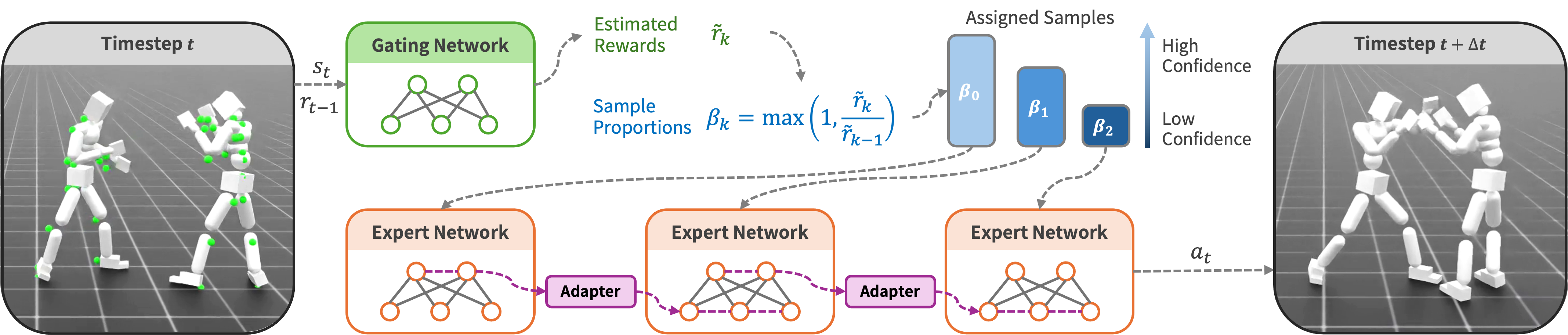}
  \caption{Framework overview. We train a progressive learning model in which later experts build on the knowledge from earlier experts, but specializing in more challenging motions. The policy receives humanoid state and goal state, and outputs actions for the proportional derivative (PD) controller to generate torques. Experts are activated sequentially, with adapters enabling knowledge transfer and a gating network estimating confidence.}
  \label{fig:architecture}
\end{figure*}

\section{Method}

\subsection{Motion Tracking}
We aim to achieve physics-based motion tracking through training a policy $\pi$ which enable a simulated humanoid character to produce a pose that closely resembles a kinematic target pose. The policy network is commonly conditioned on the future pose as a target for policy to imitate.
At each timestep $t$, with observed state $s_t$ and goal $g_t$, an RL agent interacts with an environment by applying an action $a_t$ sampled from policy $a_t \sim \pi(a_t|s_t,g_t)$ and receives a reward $r_t$. The physics simulation environment defines transition dynamics $p(s_{t+1}|s_t,a_t)$ that produces the next state $s_{t+1}$.
Similar to prior goal-conditioned reinforcement learning, we use the proximal policy gradient (PPO) to train the policy. The objective is to learn a policy that maximizes the discounted cumulative reward: 
\begin{equation}
    \mathbb{E}_{p(s_0) \prod_{t=0}^{T-1} p(s_{t+1} \mid s_t, a_t)\, \pi(a_t \mid s_t, g_t)} \left[ \sum_{t=0}^{T} \gamma_t r_t \right],
\end{equation}
where $\gamma$ is the discount factor that reduces the weight of future rewards in PPO.

\subsubsection{Observation}
The observation input consists of humanoid state $s_t$ and the goal state $g_{t+1}$ that describes the target pose for the policy to imitate. 
Humanoid state $s_t=(s^\text{p}_t, s^\text{v}_t)$ contains 
local body positions $s^\text{p}_t$ and 
linear velocities $s^\text{v}_t$. 

Goal state $g_{t+1}=(\Delta{s}^\text{p}_{t+1}, \ominus{s}^\text{q}_{t+1}, \Delta{s}^\text{v}_{t+1}, \Delta{s}^\text{a}_{t+1}, s^\text{jp}_{t+1}, s^\text{jv}_{t+1})$ contains 
body positional offset $\Delta{s}^\text{p}_{t+1}$, 
rotational difference $\ominus{s}^\text{q}_{t+1}$,
linear velocity offset $\Delta{s}^\text{v}_{t+1}$, 
angular velocity offset $\Delta{s}^\text{ja}_{t+1}$,
joint position $s^\text{jp}_{t+1}$ 
and joint velocity $s^\text{jv}_{t+1}$.

\subsubsection{Reward Function}
The reward $r_t$ encourages the agent to track the reference motion by minimizing the difference between the state of the simulated character and the ground truth:
\begin{equation}
    r^\text{track}_t = w^\text{p}r^\text{p}_t + w^\text{q}r^\text{q}_t + w^\text{v}r^\text{v}_t + w^\text{a}r^\text{a}_t + w^\text{jp}r^\text{jp}_t + w^\text{jv}r^\text{jv}_t ,
\end{equation}
where $w^{\{ \cdot \}}$ denotes the respective weights, $r^{\{ \cdot \}}$ denote the reward functions for tracking target body position, quaternion, linear velocity, angular velocity, joint position and joint velocity.

We also apply an energy penalty reward $r^\text{energy}_t$ to minimize the high-frequency jitter.
Total reward is calculated as:
\begin{equation}
    r_t = r^\text{track}_t + 0.5\times r^\text{energy}_t.
\end{equation}
Detailed calculation and parameters of rewards can be found in supplementary material.

\subsubsection{Action}
Similar to prior work \cite{peng2022ase,luo2023perpetual,tessler2024maskedmimic}, our policy generates the action $a_t\in \mathbb{R}^J$ which serves as the target for proportional derivative (PD) controller to apply torque at each joint. The action $a_t$ is sampled from a multi-dimensional Gaussian distribution $a_t \sim \mathcal{N}(\bar{a}_t, \sigma)$, where $\bar{a}_t$ is the mean action predicted by policy and $\sigma \in \mathbb{R}^J$ is learnable standard deviation.

\subsubsection{Motion Sampling Strategy}
To encourage sampling more challenging motions during policy training, we record the tracking rewards and adjust the sampling probabilities of different motion clips based on their recent tracking performance:
\begin{equation}
    p_{m,t} = \text{Softmax}(\frac{-\bar{r}^{\text{track}}_{m}}{\mathcal{T}}) ,
\end{equation}
where $\mathcal{T}$ is the annealing temperature, $\bar{r}^{\text{track}}_{m}$ is the recent average tracking reward of motion clip $m$ and $p_{m,t}$ is the calculated sampling probability of motion clip $m$ at timestep $t$.

\subsection{Progressive Interaction Tracking}
Training a single policy network for motion tracking on a large dataset could be difficult as it easily leads to catastrophic forgetting. Inspired by the continual learning paradigm of PNN, we introduce a model designed for tracking large-scale motion dataset automatically without hand-crafted dataset schedule of various difficulties.


\subsubsection{Policy Model}
Unlike vanilla PNNs that switch distinct experts for distinct tasks, our PNN experts operate additively. Each new expert does not replace the previous one but learns a residual action offset to correct the errors of the earlier frozen experts. The Gating Network is not a standard mixture-weight generator. Instead, it acts as a reward predictor that estimates how "confident" the experts are for a given state. The connection between them is governed by the routing ratio (Equation \ref{eq:sampling_ratio}). Samples with low predicted rewards (low confidence) are routed to the new expert to learn the necessary corrective offsets.

In the context of this work, "difficulty" here specifically correlates to samples with stronger physical contact forces. A key contribution of our work is that we do not manually label difficulty. Instead, the system quantitatively defines "difficult" samples as those with low estimated tracking rewards. These are the samples where previous experts fail to predict the correct joint torques required to maintain the pose against external perturbations.

Specifically, the entire policy model $\pi$ contains $k$ expert networks $\pi = (\pi_{0}, \dots , \pi_{k})$ where each expert network is a 3-layer MLP with LeakyReLU activation.
To facilitate efficient multitasking and prevent catastrophic forgetting, the architecture incorporates a gating network $f_g$ and multiple lateral adapters \cite{rusu2016progressive} that connect sequential experts. These adapters function as knowledge transfer mechanisms, allowing newly activated experts to leverage the structural embeddings of previous frozen experts to learn residual action offsets.
Instead of generating weights to blend expert outputs, the gating network is used for predicting the reward based on the input observation. 
Given an observation, gating network $f_g(s_t,g_t): \mathbb{R}^d \rightarrow \mathbb{R}^k$ outputs an independent confidence of the experts: 
\begin{equation}
    \tilde{r}_k(s_t, g_t) = \text{Sigmoid}(f_g(s_t,g_t)_k) ,
\end{equation}
where the $\text{Sigmoid}$ activation maps the output to the unit interval $[0, 1]$, ensuring that the predicted confidence $\tilde{r}_k$ is numerically consistent with the environment's normalized tracking rewards.

We start by training the first expert $\pi_0$ on the full dataset and all other experts are frozen. When the growth of estimated reward $\tilde{r}_0$ is stagnated, we stop updating the expert and activate a new one $\pi_1$ for learning harder motions by predicting the complementary offset actions on top of previous experts:
\begin{equation}
    \pi_k(s_t, g_t) = \sum_{i<k}\pi_i(s_t, g_t) .
\end{equation}

However, it is difficult to seamlessly transition to a new expert. On one hand, the new expert needs to model the action distribution from scratch, which would lead to a significant drop in reward and time-consuming re-training. On the other hand, copying the parameters from previous expert to new one would inherit the bias and prevent expert from learning new knowledge. 

We deploy two mechanisms to ensure stable transition from old to new expert while enabling the new one to learn novel knowledge effectively. 
First, we copy the parameters to all but the final layer, while zero-initializing the last layer and randomly initialize the adapters connected to the last layer (which is different from the vanilla PNN). This is helpful to keep embedding capability inherited from old expert and ensure capacity for learning to adapt prior knowledge. 
Second, we propose a \textit{Progressive Sampling Strategy} where the number of samples routed to the new expert depends on the estimated rewards compared to previous ones:
\begin{equation}
    \beta_k = \max(1, \frac{\tilde{r}_{k}(s_t, g_t)}{\tilde{r}_{k-1}(s_t, g_t)}) ,
    \label{eq:sampling_ratio}
\end{equation}
where $\beta_k$ denotes the proportion of the samples with the lowest estimated rewards that are routed to expert $\pi_k$. This strategy encourages the new expert to prioritize learning harder motions. As the new expert improves, it is gradually exposed to samples in which the previous experts are more confident, until its performance saturates and a new expert is activated again.

\begin{figure*}[tbp]
  \centering
  \includegraphics[width=\linewidth]{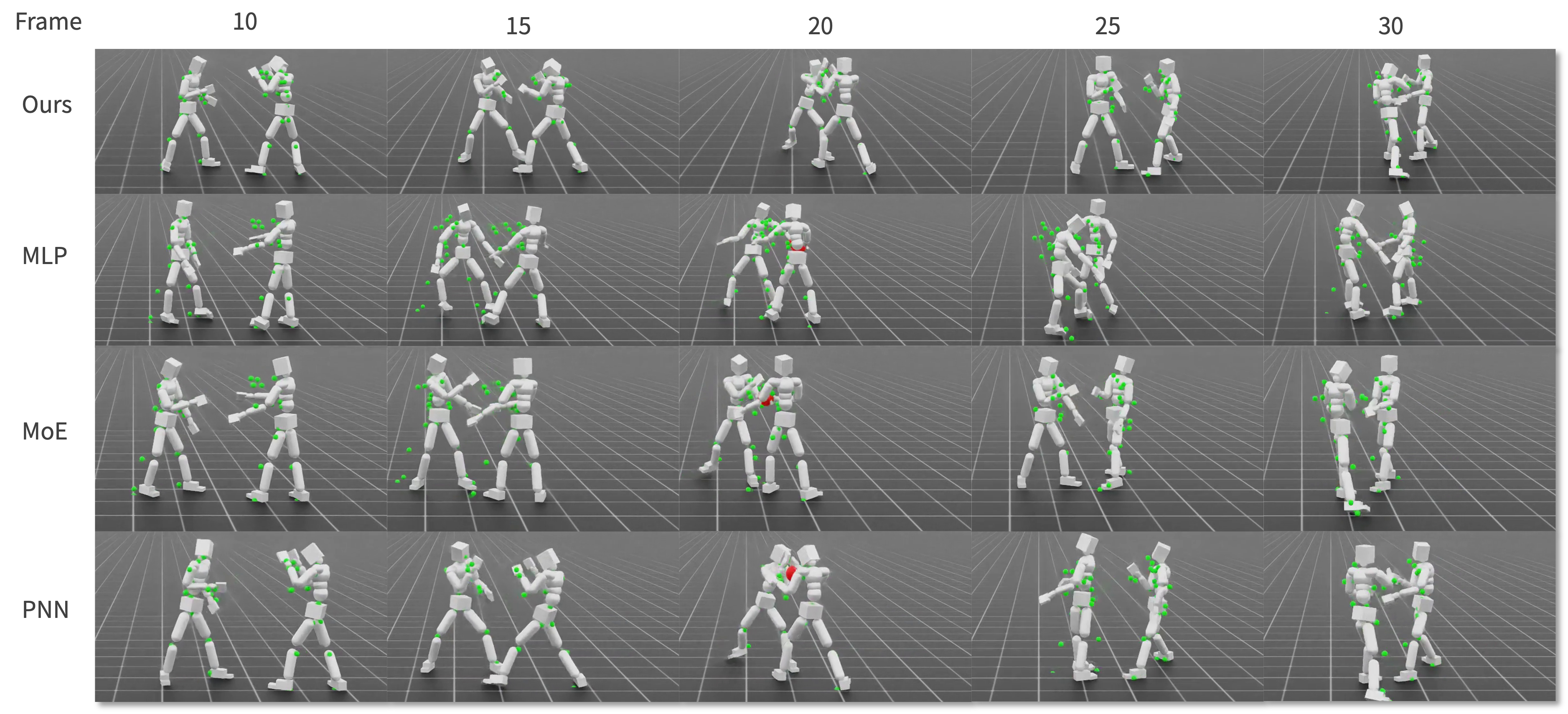}
  \caption{Qualitative comparison of tracking results across different models. From top to bottom, our method, MLP, MoE and PNN are shown that performs boxing interaction, respectively. Baseline models often exhibit instability or loss of balance under dense contact, while our method produces more stable and realistic interactions that closely follow the target motions.}
  \label{fig:comparison_tracking}
\end{figure*}

\subsubsection{Loss Function}
During training, we activate only a single expert at a time and each expert $k$ maintains its own learnable log standard deviation parameter $\log{\sigma_k}$. Using multiple $\log{\sigma_k}$ simultaneously in PPO causes unintended gradient updates in inactive experts. For the policy loss computation, we mask out the log standard deviations of all frozen experts, such that only the active expert’s $\log{\sigma_{k^*}}$ is used when constructing the Gaussian distribution:
\begin{equation}
    \pi_{k^*}(a_t | s_t) = \mathcal{N}(\bar{a}_t, \, \exp(2 \log \sigma_{k^*})).
\end{equation}
This ensures that gradients flow exclusively through the active expert, while the log stds of inactive experts remain unchanged.

In order to encourage experts to reuse the learned knowledge through adapters, we also add an adpater usage loss:
\begin{equation}
    u = \frac{\sum_{i\in \mathcal{A}_k}\lVert \theta^{\text{adapter}}_i \rVert}{\sum_{i\in \mathcal{A}_k}\lVert \theta^{\text{adapter}}_i\rVert + \lVert \theta^{\text{expert}}_{k^*}\rVert},
\end{equation}
\begin{equation}
    \mathcal{L}_{\text{adapter}} = - \log(u + \varepsilon),
    \label{eq:adapter_loss}
\end{equation}
where $\theta_{\{ \cdot \}}$ is the weights of adapters or linear layer of expert, $\mathcal{A}_k$ is the set of active adapters connecting to expert $k$ and $\varepsilon = 10^{-6}$.

For simplicity, we omit the detailed derivations of the policy and value losses, as they follow the standard PPO formulation. Total loss function is formed as $\mathcal{L} = \mathcal{L}_{\text{policy}} + \mathcal{L}_{\text{value}} + 0.03\times \mathcal{L}_{\text{adapter}}$.

For the gating network, it is updated independently from the tracking policy using the actual rewards from the environment:
\begin{equation}
    \mathcal{L}_{\text{gating}} = \sqrt{\sum_k(r_k - \tilde{r}_k)^2} .
\end{equation}

\section{Experiments}
\subsection{Implementation Details}
We conduct our experiments with an NVIDIA A100 GPU for training 4 experts. The physics simulation is run on NVIDIA Isaac Lab \cite{mittal2023orbit}. We use SMPL \cite{loper2023smpl} kinematic structure for the humanoid which contain 24 rigid bodies (pelvis as the root joint) and 69 degrees of freedom. 
Following prior work \cite{peng2018deepmimic}, the starting frame of training episode is randomly selected from the sampled motion clip and the episode will be early terminated if the averaged rigid body distance to the target is lower than 0.5m.
Our model is trained on InterHuman \cite{liang2024intergen} which contains motion sequences of 1 million frames (10 hours at 30 fps).

\subsection{Baselines}
We compare our method against three baseline tracking policy models: 
\begin{itemize}
    \item A 3-layer multi-layer perceptron network (MLP) that represents the basic policy implementation such as SONIC \cite{luo2025sonic}.
    \item Mixture-of-experts network (MoE) which is representative for learning diverse skills used in previous works like DeepMimic \cite{peng2018deepmimic}.
    \item Progressive Neural Network (PNN) \cite{rusu2016progressive} that represents the manual progressive strategies used in recent SOTA like PHC \cite{luo2023perpetual}. PHC relies on a manually scheduled progressive network; our PNN baseline implements this exact manual strategy to demonstrate its limitations compared to our automatic routing. Specifically, the full dataset is partitioned into four subsets based on the averaged relative distances between the two humanoid robots, with later experts assigned to motions involving more frequent inter-body contacts.
\end{itemize}

\subsection{Results}
\subsubsection{Tracking Performance}
Following prior work \cite{luo2023perpetual,tessler2024maskedmimic}, in Table \ref{tab:compare_tracking}, we evaluate the tracking success rate (referred to as `Success') as the ratio of successful episodes in which the average joint position error at every frame is less than 0.5m. We also report the mean per-joint position error (MPJPE) to assess the accuracy of alignment with the target pose. Despite the frequent perturbations of contact, our approach still achieves robust performance and a significantly higher success rate.

\begin{figure}[tbp]
  \centering
  \includegraphics[width=\linewidth]{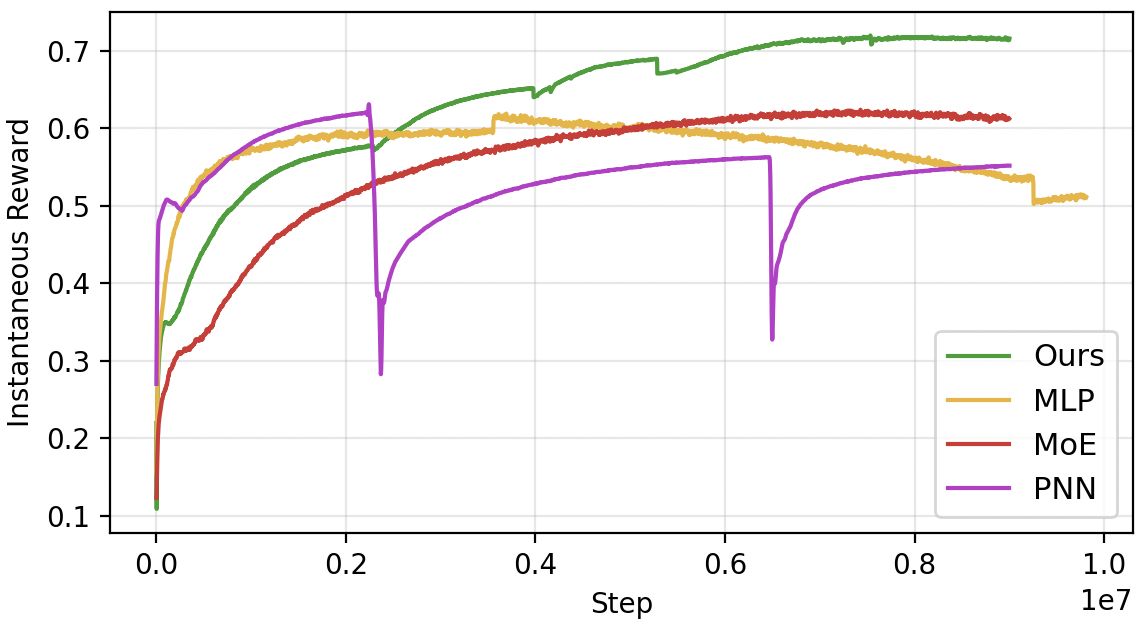}
  \caption{Training reward curves of baselines and our method. Our approach shows smooth transitions between experts and faster convergence compared to PNN.}
  \label{fig:rewards}
\end{figure}

Figure \ref{fig:rewards} shows the tracking reward curves of all methods. MLP quickly saturates in the early stage and later degrades due to catastrophic forgetting. MoE achieves higher rewards than MLP with its larger capacity but still suffers from forgetting. PNN exhibits sharp drops when new experts are activated, caused by the distribution shift across motion subsets, and also incurs higher training cost since each expert must be trained from scratch. 
In contrast, our method exhibits substantially more stable transitions when introducing new experts and requires considerably less training time. Moreover, it automatically routes appropriate number of samples to the active experts, thereby obviating the need for manual dataset scheduling across experts in progressive learning.

\begin{figure*}[tbp]
  \centering
  \includegraphics[width=\linewidth]{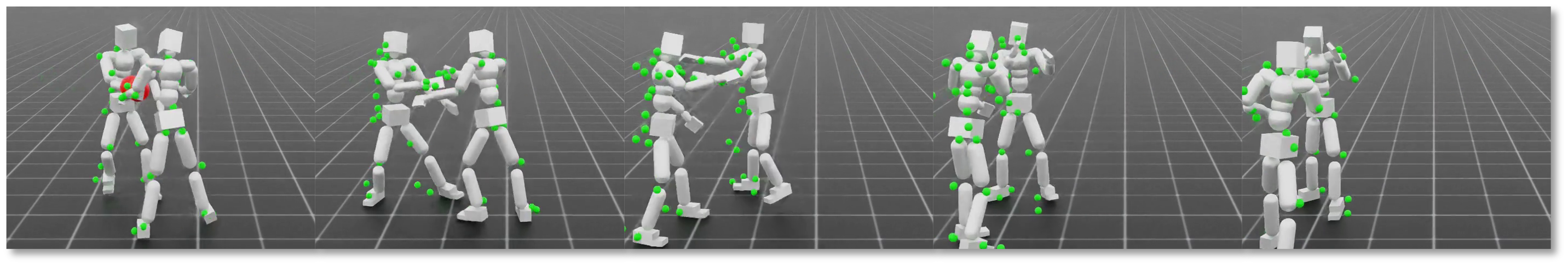}
  \caption{Tracking with interaction skill transitions. The target interaction shifts from spinning to boxing, and our system remains robust to these abrupt changes without failure.}
  \label{fig:skill_transition}
\end{figure*}

As illustrated in Figure \ref{fig:skill_transition}, our method also demonstrates strong adaptability when the interaction skill transitions abruptly from spinning to boxing. Despite the sudden change in motion dynamics, the characters maintain stable coordination without collapsing into unnatural states. This indicates that the progressive expert routing effectively preserves prior knowledge while enabling quick adaptation to new interaction modes. Notably, the system avoids discontinuous prediction across multiple experts, which are common in skill-specific controllers such as PNN.

\begin{table}[!ht]
    \centering
    \begin{tabular}{l|cc|cc}
        ~ & \multicolumn{2}{c|}{Train Set} & \multicolumn{2}{c}{Test Set} \\
        ~ & Success$\uparrow$ & MPJPE$\downarrow$ & Success$\uparrow$ & MPJPE$\downarrow$ \\
        \hline
        MLP & 53.3\% & 81.3 & 28.5\% & 117.0 \\
        MoE & 76.2\% & 55.7 & 64.0\% & 69.4 \\ 
        PNN & 82.3\% & 50.9 & 78.7\% & 58.1 \\ 
        Ours & \textbf{91.8\%} & \textbf{37.9} & \textbf{85.2\%} & \textbf{43.4} \\ 
    \end{tabular}
    \caption{Comparison of tracking performance across baselines and our method on InterHuman dataset. We report success rate and mean per-joint position error (MPJPE) on both training and test sets.}
    \label{tab:compare_tracking}
\end{table}

We further evaluate our method on a subset of the AMASS dataset to examine its performance in single-character tracking scenarios. This subset consists of approximately 200 motion clips, totaling 300 minutes of motion sequences at 30 fps. For the PNN baseline, the dataset is partitioned into four subsets assigned to four corresponding experts based on averaged joint velocities. As shown in Table \ref{tab:compare_AMASS}, all models demonstrate higher tracking success rates on AMASS compared to the InterHuman dataset, which is primarily attributed to the absence of inter-body perturbations inherent in two-character interactions.
\begin{table}[!ht]
    \centering
    \begin{tabular}{l|cc|cc}
        ~ & \multicolumn{2}{c|}{Train Set} & \multicolumn{2}{c}{Test Set} \\
        ~ & Success$\uparrow$ & MPJPE$\downarrow$ & Success$\uparrow$ & MPJPE$\downarrow$ \\
        \hline
        MLP & 71.5\% & 56.0 & 66.5\% & 72.3 \\
        MoE & 86.3\% & 47.2 & 82.9\% & 53.1 \\ 
        PNN & 91.6\% & 42.8 & 88.4\% & 46.5 \\ 
        Ours & \textbf{96.7\%} & \textbf{36.1} & \textbf{93.0\%} & \textbf{40.2} \\ 
    \end{tabular}
    \caption{Comparison of tracking performance across baselines and our method on AMASS dataset.}
    \label{tab:compare_AMASS}
\end{table}

\begin{figure}[tbp]
  \centering
  \includegraphics[width=\linewidth]{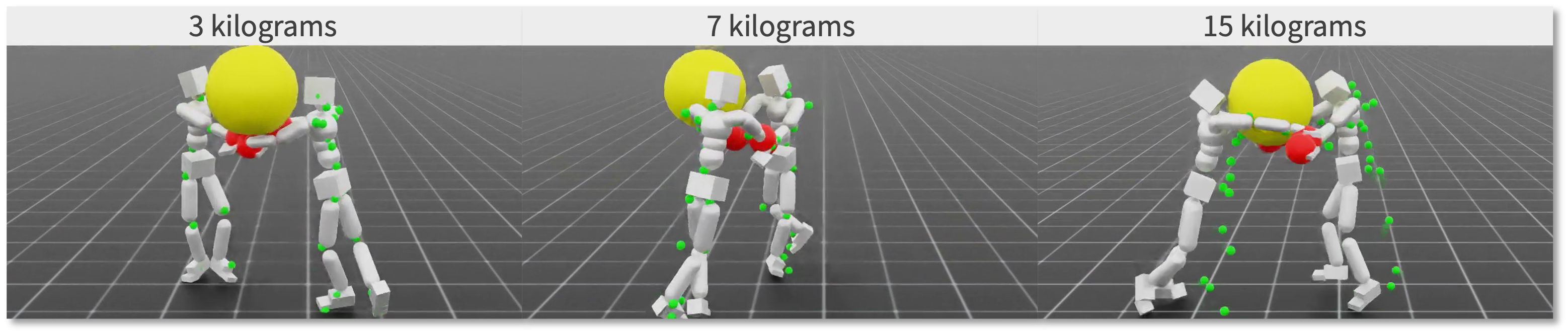}
  \caption{Tracking under external perturbations with different object masses (3 kg, 7 kg, 15 kg). As the perturbation strength increases, the characters experience growing difficulty in maintaining stable interaction.}
  \label{fig:comparison_object_perturb_kg}
\end{figure}

\subsubsection{Perturbations}
To assess robustness, we introduce perturbations to the tracked humanoid by randomly throwing objects of varying masses and by injecting noise into the input observations. 
In these experiments, all models, including the baselines, are retrained in environments where such perturbations are applied during training. Quantitative and qualitative results are reported in Table \ref{tab:compare_perturbations}. 

When tested with external disturbances, our approach sustains high success rates, while baseline methods show marked degradation. 
Notably, PNN fails under perturbations due to its gating network’s reliance on dataset-specific specialization, which does not generalize when noise shifts the input distribution. 
MoE handles perturbations slightly better through soft blending, but still lacks sufficient adaptability. Since MoE and MLP are already prone to catastrophic forgetting, they perform even worse in noisy environments compared to safe ones.
Our experts, trained progressively with sample routing, exhibit significantly stronger resilience to observation noise and external force perturbations. This suggests that expert specialization in our framework is not brittle but rather complementary, where later experts refine challenging behaviors without overwriting earlier skills.

\begin{table}[!ht]
    \centering
    \begin{tabular}{l|cc|cc}
        ~ & \multicolumn{2}{c|}{Object Perturb} & \multicolumn{2}{c}{Noise Inject} \\
        ~ & Success$\uparrow$ & MPJPE$\downarrow$ & Success$\uparrow$ & MPJPE$\downarrow$ \\
        \hline
        MLP & 14.5\% & 177.6 & 20.3\% & 139.9 \\
        MoE & 52.4\% & 90.6 & 57.1\% & 81.1 \\ 
        PNN & 55.3\% & 84.3 & 59.5\% & 77.0 \\ 
        Ours & \textbf{75.3\%} & \textbf{59.0} & \textbf{80.2\%} & \textbf{52.9} \\ 
    \end{tabular}
    \caption{Robustness evaluation under perturbations. Performance is measured when randomly throwing objects of varying masses (Object Perturb) and injecting noise into input observations (Noise Inject).}
    \label{tab:compare_perturbations}
\end{table}

\begin{figure}[tbp]
  \centering
  \includegraphics[width=\linewidth]{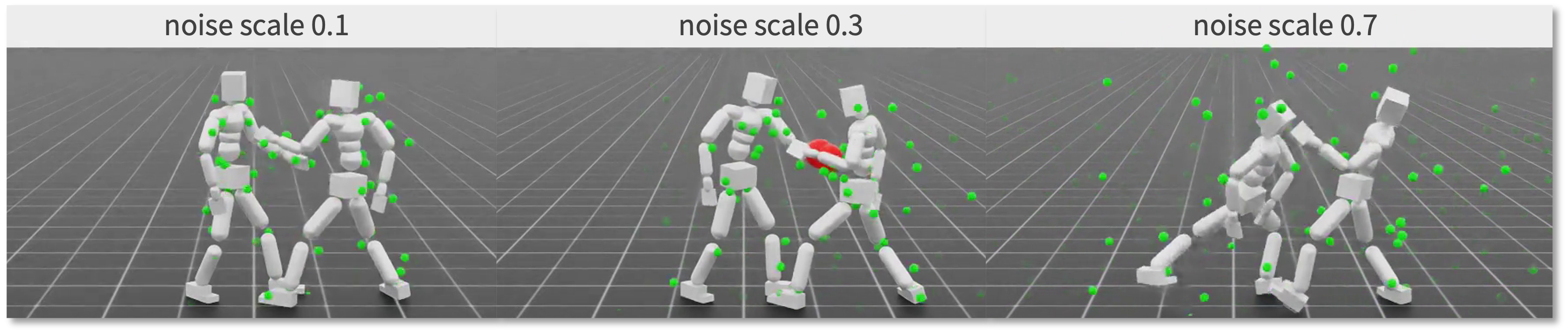}
  \caption{Tracking under observation noise with different noise scales (0.1, 0.3, 0.7). Larger noise levels lead to instability and loss of balance in the interactions.}
  \label{fig:comparison_noise_inject}
\end{figure}

\subsubsection{Expert Contribution}
In our method, later experts enhance learning capacity by specializing in more challenging input samples and predicting action offsets for earlier experts.
The results in Table \ref{tab:compare_experts} sheds light on how different experts contribute during training. We observe that each newly activated expert reduces both MPJPE and training time, which indicates that later experts act as refiners by predicting action offsets relative to earlier experts. Interestingly, the marginal improvement between the fourth and fifth expert is small, suggesting diminishing returns once the model capacity surpasses the scale of InterHuman. This analysis confirms that four experts are sufficient for the present dataset, and that our automatic routing strategy naturally balances capacity and efficiency without requiring manual dataset scheduling.more quickly.

Another important observation is that our approach substantially reduces training time compared to PNN. In PNN, each expert is trained nearly from scratch on a subset of data, resulting in redundant training and sharp reward drops. Our method mitigates this by transferring knowledge across experts, as shown by the smooth reward curve (Figure \ref{fig:rewards}). This not only improves efficiency but also avoids instability during expert activation, which is crucial for scaling to larger datasets.

\begin{table}[!ht]
    \centering
    \begin{tabular}{l|c|c|c}
        ~ & Success$\uparrow$ & MPJPE$\downarrow$ & Training Time (hours)\\
        \hline
        Expert 1 & 61.1\% & 83.4 & 29 \\
        Expert 2 & 72.6\% & 60.2 & 19\\ 
        Expert 3 & 82.7\% & 47.5 & 15 \\ 
        Expert 4 & 85.2\% & 43.4 & 7 \\ 
        Expert 5 & 85.8\% & 42.7 & 2 \\ 
    \end{tabular}
    \caption{Contribution of different experts in our progressive framework. Later experts improve success rate and MPJPE while requiring less training time, showing that specialization accelerates learning.}
    \label{tab:compare_experts}
\end{table}

\begin{figure*}[tbp]
  \centering
  \includegraphics[width=\linewidth]{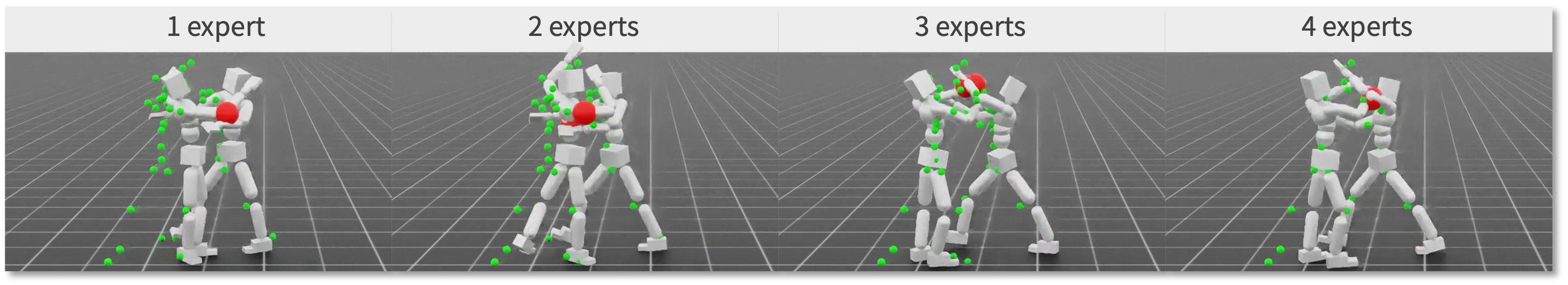}
  \caption{Ablation on the number of experts in our progressive framework. With only 1–2 experts, the characters often fail to maintain stable interactions. Adding more experts (3–4) improves tracking quality, showing that later experts specialize in handling more challenging motion dynamics.}
  \label{fig:comparison_experts}
\end{figure*}

\subsubsection{Ablation Study}
We conduct experiments ablating key components of our framework: (1) Progressive Sampling Strategy (PSS), where all new experts are trained on the full dataset when initiated instead of routing based on estimated rewards; (2) adapters, removing the lateral connections between experts that enable knowledge transfer; and (3) adapter loss, omitting the regularization term that encourages adapter usage (Equation \ref{eq:adapter_loss}).

\begin{table}[!ht]
\footnotesize
    \centering
    \begin{tabular}{l|c|c|c}
        ~ & Success$\uparrow$ & MPJPE$\downarrow$ & Training Time (hours)\\
        \hline
        Full Setup & \textbf{85.2\%} & \textbf{43.4} & 70 \\ 
        w/o PSS & 68.0\% & 64.1 & 167\\ 
        w/o adapter & 76.3\% & 60.9 & 122 \\ 
        w/o adapter loss & 82.3\% & 49.1 & 70 \\ 
    \end{tabular}
    \caption{Qualitative results of full setup and ablated versions of our method.}
    \label{tab:ablation}
\end{table}

Removing Progressive Sampling Strategy causes the largest performance drop and substantially increases training time. This reveals that the model wastes capacity re-learning easy samples for each expert instead of specializing on hard, contact-heavy failures. 
Ablating the adapters also degrades tracking quality and slows training. This shows that progressive experts cannot function in isolation. Without lateral knowledge transfer, each new expert is forced to re-learn large parts of the representation space, leading to redundant computation
Removing the adapter loss leads to a smaller but noticeable decline in performance. The loss acts mainly as a fine-tuning mechanism rather than a structural necessity. It gently encourages later experts to reuse previous knowledge via adapters instead of over-specializing or ignoring them.

\subsubsection{Motion Sampling Strategy}
We further evaluate the impact of different motion sampling strategies on performance. Specifically, we compare four approaches: (1) \textit{Uniform}, where all motion clips are sampled with equal probability; (2) \textit{Motion Duration}, where longer clips are assigned higher sampling probabilities; (3) \textit{Success Rate}, where clips with lower success rates are prioritized to encourage training on more difficult motions; and (4) \textit{Tracking Reward}, where clips with lower tracking rewards are sampled more frequently to emphasize challenging cases. 
In Table \ref{tab:compare_sampling_strategy}, we report both the success rate and the mean episode length (referred to as `Episode') in training stage. The mean episode length reflects the average duration of simulation an agent survives before the episode ends, either due to task termination (e.g., falling or large tracking error) or truncation at the maximum allowed horizon.

Table \ref{tab:compare_sampling_strategy} highlights how different motion sampling strategies influence tracking performance. The Uniform strategy yields the lowest success rate and episode length, as it does not differentiate between motions of varying complexity. 
Motion Duration provides a modest improvement, suggesting that longer clips contain more diverse patterns that benefit learning. However, the gains remain limited because this strategy does not explicitly prioritize difficult motions.

In contrast, Success Rate sampling substantially improves both success and episode length. By down-weighting easier motions and emphasizing those with lower success, the policy is exposed to more challenging interactions. 
Similarly, Tracking Reward achieves the best overall performance. Prioritizing motions with lower tracking reward ensures that the policy repeatedly trains on failure-prone sequences, which leads to longer survival time during episodes.
Interestingly, the difference between Success Rate and Tracking Reward is relatively small, but the latter consistently achieves the highest values. This suggests that reward-based sampling provides a finer-grained signal of difficulty compared to binary success or failure.

\begin{table}[!ht]
    \centering
    \begin{tabular}{l|c|c}
        ~ & Success$\uparrow$ & Episode (seconds)$\uparrow$ \\
        \hline
        Uniform & 74.4\% & 14.9 \\
        Motion Duration & 81.7\% & 17.1 \\ 
        Success Rate & 88.4\% & 20.5 \\ 
        Tracking Reward & \textbf{91.8\%} & \textbf{21.2} \\ 
    \end{tabular}
    \caption{Impact of motion sampling strategies. Adaptive sampling based on success rate or tracking reward leads to higher success and longer average episode length compared to uniform or duration-based sampling.}
    \label{tab:compare_sampling_strategy}
\end{table}

\section{Conclusion}
We introduced a progressive mixture-of-experts framework for physics-based motion tracking of contact-rich interactions. By progressively expanding expert capacity and automatically routing training samples, our method achieves stable and efficient learning without requiring manual dataset scheduling. 
Experiments on large-scale datasets demonstrate that our approach surpasses prior baselines in tracking accuracy, robustness under perturbations, and training efficiency. Ablation studies further confirm that later experts specialize in more challenging dynamics while maintaining smooth knowledge transfer across the model. 
We believe this all-in-one progressive framework offers an extensible foundation for the community to advance research in motion tracking without reliance on task-specific controllers or fragmented training pipelines.

\section*{Acknowledgment}
This project is supported in part by the EPSRC NortHFutures project (ref: EP/X031012/1).

\printbibliography                

\end{document}